\documentclass[conference]{IEEEtran}
\usepackage{amsmath}
\usepackage{array}
\usepackage{amssymb,amsmath,color,graphicx,amsbsy,bm}
\usepackage{graphicx}
\usepackage{subfigure}
\usepackage{cite}
\usepackage{algorithm}
\usepackage{algorithmic}
\usepackage{amsmath}
\usepackage{amsfonts}
\usepackage{amssymb}
\usepackage{graphicx}
\usepackage{multirow}
\allowdisplaybreaks
\usepackage{soul}

\begin{document}
%
\title{A DSO Framework for Comprehensive Market Participation of DER Aggregators}

\author{\IEEEauthorblockN{Mohammad Mousavi, \textit{Student Member, IEEE}}
\IEEEauthorblockA{Arizona State University\\
Email: mmousav1@asu.edu}
\and
\IEEEauthorblockN{Meng Wu, \textit{Member, IEEE}}
\IEEEauthorblockA{Arizona State University\\
Email: mwu@asu.edu}
}


%


\maketitle

\begin{abstract}
In this paper, a distribution system operator (DSO) framework is proposed to optimally coordinate distributed energy resources (DER) aggregators' comprehensive participation in retail energy market as well as wholesale energy and regulation markets. Various types of DER aggregators, including energy storage aggregators (ESAGs), dispatchable distributed generation aggregators (DDGAGs), electric vehicles charging stations (EVCSs), and demand response aggregators (DRAGs), are modeled in the proposed DSO framework. Distribution network constraints are considered by using a linearized power flow. The problem is modeled using mixed-integer linear programming (MILP) which can be solved by commercial solvers. Case studies are performed to analyze the interactions between DER aggregators and wholesale/retail electricity markets.

\end{abstract}


%
\IEEEpeerreviewmaketitle

\section{Introduction}
Due to environmental issues and increasing demand, the installed capacity of distributed energy resources (DER) is growing rapidly. DER aggregators, with low operating costs and fast ramping capability, can effectively participate in the wholesale energy and regulation markets. However, to participate in the wholesale markets, DER aggregators need to control DER power outputs across the distribution network, which will cause security and reliability issues to the distribution system operation. Hence, there is a need for an entity that coordinate DER aggregators to participate in the wholesale and retail markets while ensuring distribution network security.

Recently, many issues have been investigated for DER market participation\cite{DiSomma,Baringo,Anjos,Liu,Lezama,Nguyen,Yang}. In \cite{DiSomma}, the DER aggregator is defined to enable DER market participation. In \cite{Baringo}, DER wholesale market participation is enabled through the virtual power plant. In \cite{Anjos}, a decentralized approach, based on Dantzig-Wolfe decomposition, is proposed for DER coordination. This approach allows a numerous number of households to interact with an aggregator to minimize the total cost of purchasing electricity. In \cite{Liu,Lezama}, the optimal operation of a microgrid for its wholesale market participation is presented. Above previous works neglect the distribution network power flow constraints, therefore ignore the distribution network security while coordinating DER market participation. In \cite{Nguyen}, the bidding strategy of the virtual power plant considering the demand response market is presented. The demand response market is defined as a stage between the day-ahead market and the real-time market. In \cite{Yang}, the optimal bidding strategy of EV aggregators for participating in the day-ahead and the real-time markets is presented. In \cite{Nguyen,Yang}, DC power flow is presented as distribution power balance constraints, which is inappropriate due to high impedances in distribution grids.

Motivated by the increasing DER penetration level and emerging smart distribution grid technologies, the power industry calls for a distirbution operation framework which can handle DER market participation at the distribution level while respecting the distirbution system operating constraints. Recently, the distribution system operator (DSO) is introduced to operate the distribution system and retail market with DER integration \cite{Faqiry,doPrado2,Parhizi}. In \cite{Faqiry}, a day-ahead market framework operated by a DSO is presented. The DSO pays the distribution market participants at distribution locational marginal prices (D-LMPs). However, the distribution network and related constraints are not considered in the proposed model. In \cite{doPrado2}, a two-stage stochastic programming is applied to model day-ahead energy and reserve markets operated by a DSO. In \cite{Parhizi}, a distribution market operator (DMO) is defined which gathers offers from microgrids and aggregates them to participate in the wholesale market. A penalty factor is defined to reprensent the relationship betwen D-LMP and transmission-level LMP. Both \cite{doPrado2} and \cite{Parhizi} adopt DC power flow as the distribution system model, which is insufficient as discussed previously.

To the best of our knowledge, the DSO framework for optimal coordination of DER aggregators' participation in wholesale energy and regulation markets as well as retail energy market has not been studied yet. In this paper, a DSO framework is proposed based on the mixed-integer linear programming (MILP) formulation. The proposed DSO operates the reatil energy market and also gathers offers from DER aggregators for wholesale energy and regulation markets participation. Various types of DER aggregators, including energy storage aggregators (ESAGs), dispatchable distributed generation (DG) aggregators (DDGAGs), electric vehicles (EV) charging stations (EVCSs), and demand response aggregators (DRAGs), are considered in the proposed DSO framework. Moreover, the distribution network constraints are considered using a linearized power flow. Case studies are performed to analyze the interactions between DER aggregators and wholesale/retail electricity markets.
   
\section{DSO Market Formulation}
In this paper, the DSO is defined as a mediator that participates in the wholesale markets on one side and interacts with DER aggregators and end-user customers on the other side. Various types of DER aggregators submit their offers to the DSO. The DSO collects these offers to operate the retail market as well as coordinate the offers to construct an aggregated bid for participating in day-ahead wholesale energy and regulation markets operated by the independent system operator (ISO). At the wholesale level, this paper assumes the market framework of California ISO (CAISO), whose pay-for-performance regulation market considers offers for both regulation capacity (with capacity-up and capacity-down offers) and regulation mileage \cite{CAISO}. The DSO is modeled as a price-taker in the day-ahead wholesale market. The MILP formulation of this DSO framework is presented below.

\subsection{Objective Function}	
The proposed DSO minimizes the total operating cost while maximizing total social welfare in the distribution network. The regulation market model in \cite{Fooladivanda,Vatandoust} is adopted. The objective function is presented in (\ref{equ.1}).\\
\begin{equation}\label{equ.1}
\begin{split}
min &\sum_{t \in T}[ -P^{sub}_{t}\pi^{e}_{t}-r^{sub,up}_{t}\pi^{cap,up}_{t}-r^{sub,dn}_{t}\pi^{cap,dn}_{t}\\
&-r^{sub,up}_{t}S^{up}_{t}\mu^{up}_{t}\pi^{mil,up}_{t}-r^{sub,dn}_{t}S^{dn}_{t}\mu^{dn}_{t}\pi^{mil,dn}_{t}\\
&+\sum_{k \in\{K_{2},K_{4}\}}P_{t,k}\pi^{e}_{t,k}-\sum_{k_{3} \in K_{3}}P_{t,k_{3}}\pi^{e}_{t,k_{3}}\\
&+\sum_{k \in K}[ r^{up}_{t,k}\pi^{cap,up}_{t,k}+r^{dn}_{t,k}\pi^{cap,dn}_{t,k}\\
&+r^{up}_{t,k}S^{up}_{t}\mu^{up}_{t}\pi^{mil,up}_{t,k}+r^{dn}_{t,k}S^{dn}_{t}\mu^{dn}_{t}\pi^{mil,dn}_{t,k}]  \\
&-\sum_{k_{1} \in K_{1}}\sum_{a \in A} P_{a,t,k_{1}}\pi^{e}_{a,t,k_{1}}] 
\end{split}
\end{equation}
where $t$ and $T$ are the index and set for the entire operating timespan; $k$ and $K=\{K_{1}, K_{2}, K_{3}, K_{4}\}$ are the index and set for all DER aggregators; $k_{1}$ and $K_{1}$ are the index and set for all DRAGs; $k_{2}$ and $K_{2}$ are the index and set for all ESAGs; $k_{3}$ and $K_{3}$ are the index and set for all EVCSs; $k_{4}$ and $K_{4}$ are the index and set for all DDGAGs; $a$ and $A$ are the index and set for all demand blocks; $P_t^{sub},r_t^{sub,up}$, and $r_t^{sub,dn}$ are the DSO's aggregated offers to wholesale energy, regulation capacity-up and capacity-down markets, respectively; $\pi_t^{e},\pi_t^{cap,up}$, and $\pi_t^{cap,dn}$ are the wholesale energy, regulation capacity-up, and capacity-down prices, respectively;  $\pi_t^{mil,up}$ and $\pi_t^{mil,dn}$ are the wholesale regulation mileage-up and mileage-down prices; $P_{t,k},r_{t,k}^{up}$ and $r_{t,k}^{dn}$ are the energy, regulation capacity-up and capacity-down offers made by DER aggregator $k$ with corresponding prices $\pi_{t,k}^{e},\pi_{t,k}^{cap,up},\pi_{t,k}^{cap,dn}$, respectively; $\mu_t^{up}$ and $\mu_t^{dn}$ are historical scores for providing regulation mileage-up and mileage-down services; $S_t^{up}$ and $S_t^{dn}$ are the regulation mileage-up and mileage-down ratios (the expected mileage for $1 MW$ provided regulation capacity); $P_{a,t,k_{1}}$ and $\pi^{e}_{a,t,k_{1}}$ are the power consumption and the corresponding energy price at each demand block.

\subsection{Constraints for Demand Response Aggregators (DRAGs)}
The operating constraints for DRAGs are as follows:
\begin{align}
& \sum_{a \in A} P_{a,t,k_{1}}-r_{t,k_{1}}^{cap,dn} \ge 0; \hspace{3mm}\forall t\in T,\, \forall k_{1} \in K_{1}\label{equ.2}\\
& \sum_{a \in A} P_{a,t,k_{1}}+r_{t,k_{1}}^{cap,up}\le \sum_{a \in A} P_{a,k1}^{max};\hspace{3mm}\forall t\in T,\, \forall k_{1} \in K_{1} \label{equ.3}\\
& 0 \le P_{a,t,k_{1}} \le P_{a,k1}^{max};\hspace{3mm}\forall a \in A,\,\forall t\in T,\, \forall k_{1} \in K_{1}\label{equ.4}\\
& 0 \le r_{t,k_{1}}^{cap,up} \le r_{t,k_{1}}^{cap,up,max};\hspace{3mm}\forall t\in T,\, \forall k_{1} \in K_{1} \label{equ.5}\\
& 0 \le r_{t,k_{1}}^{cap,dn} \le r_{t,k_{1}}^{cap,dn,max};\hspace{3mm}\forall t\in T,\, \forall k_{1} \in K_{1} \label{equ.6}
\end{align}
where $P_{a,t,k{1}}^{max}$ is the maximum power consumption at each demand block; $r_{t,k{1}}^{cap,up,max}$ and $r_{t,k{1}}^{cap,dn,max}$ are the maximum allowed regulation capacity-up and capacity-down offers.

Equations (\ref{equ.2}) and (\ref{equ.3}) limit the DRAG's offers to energy, regulation capacity-up and capcity-down markets. Equation (\ref{equ.4}) ensures that the real power offered at each demand block is within its permitted range. Equations (\ref{equ.5}) and (\ref{equ.6}) ensure that the regulation capacity-up and capacity-down offers are lower than their maximum permitted values.

\subsection{Constraints for Energy Storage Aggregators (ESAGs)}
The operating constraints for ESAGs are as follows:
\begin{align}
\begin{split}\label{equ.7}
&P_{t,k_{2}}=E_{t-1,k_{2}}-E_{t,k_{2}}+(1/\eta_{k_{2}}^{di})r_{t,k_{2}}^{cap,up}\mu_{t}^{up}\\
&\hspace{10mm}-(\eta_{k_{2}}^{ch})r_{t,k_{2}}^{cap,dn}\mu_{t}^{dn};\hspace{3mm}\forall t\in T,\, \forall k_{2} \in K_{2} 
\end{split}\\
& P_{t,k_{2}}=(1/\eta_{k_{2}}^{di})P_{t,k_{2}}^{di}-(\eta_{k_{2}}^{ch})P_{t,k_{2}}^{ch};\hspace{3mm}\forall t\in T,\, \forall k_{2} \in K_{2}\label{equ.8}\\
& r_{t,k_{2}}^{cap,up}=r_{t,k_{2}}^{cap,up,di}+r_{t,k_{2}}^{cap,dn,ch};\hspace{3mm}\forall t\in T,\, \forall k_{2} \in K_{2} \label{equ.9}\\
& r_{t,k_{2}}^{cap,dn}=r_{t,k_{2}}^{cap,dn,di}+r_{t,k_{2}}^{cap,up,ch};\hspace{3mm}\forall t\in T,\, \forall k_{2} \in K_{2} \label{equ.10}\\
& E_{k_{2}}^{min} \le E_{t,k_{2}} \le E_{k_{2}}^{max};\hspace{3mm}\forall t\in T,\, \forall k_{2}  \in K_{2}\label{equ.11}\\
& 0 \le P_{t,k_{2}}^{di} \le b_{t,k_{2}} DR_{k_{2}}^{max} ;\hspace{3mm}\forall t\in T,\, \forall k_{2} \in K_{2}\label{equ.12}\\
& 0 \le r_{t,k_{2}}^{cap,up,di} \le b_{t,k_{2}} DR_{k_{2}}^{max};\hspace{3mm}\forall t\in T,\, \forall k_{2}  \in K_{2}\label{equ.13}\\
& 0 \le r_{t,k_{2}}^{cap,dn,di} \le b_{t,k_{2}} DR_{k_{2}}^{max};\hspace{3mm}\forall t\in T,\, \forall k_{2} \in K_{2} \label{equ.14}\\
& 0 \le P_{t,k_{2}}^{ch} \le (1-b_{t,k_{2}})CR_{k_{2}}^{max};\hspace{3mm}\forall t\in T,\, \forall k_{2}  \in K_{2}\label{equ.15}\\
& 0 \le r_{t,k_{2}}^{cap,up,ch} \le (1-b_{t,k_{2}})CR_{k_{2}}^{max};\hspace{0mm}\forall t\in T, \forall k_{2}  \in K_{2}\label{equ.16}\\
& 0 \le r_{t,k_{2}}^{cap,dn,ch} \le (1-b_{t,k_{2}})CR_{k_{2}}^{max};\hspace{0mm}\forall t\in T, \forall k_{2} \in K_{2} \label{equ.17}\\
\begin{split}\label{equ.18}
& r_{t,k_{2}}^{cap,dn,di} \le P_{t,k_{2}}^{di} \le DR_{k_{2}}^{max}-r_{t,k_{2}}^{cap,up,di};\\
&\hspace{27mm}\forall t\in T,\, \forall k_{2} \in K_{2}
\end{split}\\
\begin{split}\label{equ.19}
& r_{t,k_{2}}^{cap,dn,ch} \le P_{t,k_{2}}^{ch} \le CR_{k_{2}}^{max}-r_{t,k_{2}}^{cap,up,ch};\\
&\hspace{27mm}\forall t\in T,\, \forall k_{2} \in K_{2}
\end{split}
\end{align}
where $E_{t,k_{2}}$ is the charging level; $\eta_{k_{2}}^{ch}$ and $\eta_{k_{2}}^{di}$ are the charging and discharging efficiancies; $P_{t,k_{2}}^{di}$ is the discharging power; $P_{t,k_{2}}^{ch}$ is the charging power; $r_{t,k_{2}}^{cap,up,di}$ and  $r_{t,k_{2}}^{cap,dn,di}$ are the regulation capacity-up and capacity-down offers in the discharging mode; $r_{t,k_{2}}^{cap,dn,ch}$ and $r_{t,k_{2}}^{cap,up,ch}$ are the regulation capacity-up and capacity-down offers in the charging mode; ${CR}_{k_{2}}^{max}$ and ${DR}_{k_{2}}^{max}$ are the maximum charging and discharging rates; $b_{t,k_{2}}$ is a binary variable indicating the charging ($b_{t,k_{2}}=0$) and discharging ($b_{t,k_{2}}=1$) modes.

Equation (\ref{equ.7}) represents ESAG's power injection. ESAG's offers to the energy, regulation capacity-up and capacity-down markets are decomposed into charging and discharging terms by Equations  (\ref{equ.8})-(\ref{equ.10}). Equation (\ref{equ.11}) limits the charge level of ESAGs. Equations (\ref{equ.12})-(\ref{equ.17}) ensure that ESAG's offers to the energy, regulation capacity-up and capacity-down markets are in their permitted ranges. Equations (\ref{equ.18})-(\ref{equ.19}) limit ESAG's offers to the energy, regulation capacity-up and capacity-down markets with respect to the charging and discharging rates.
\subsection{Constraints for EV Charging Stations (EVCSs)}
EVCSs are modeled as EV charging aggregators and are assumed to have unidirectional power flow as assumed in \cite{Vatandoust}. Constraints related to the operation of EVCSs are as follows: 
\begin{align}
& 0\le P_{t,k_{3}}\le ER_{k_{3}}^{max}b_{k_{3}};\hspace{3mm}\forall t\in T^{'},\, \forall k_{3} \in K_{3} \label{equ.20_1}\\
&0\le  r_{t,k_{3}}^{cap,up} \le ERR_{k_{3}}^{max}b_{k_{3}};\hspace{3mm}\forall t\in T^{'},\, \forall k_{3} \in K_{3} \label{equ.20_2}\\
&0\le  r_{t,k_{3}}^{cap,dn} \le ERR_{k_{3}}^{max}b_{k_{3}};\hspace{3mm}\forall t\in T^{'},\, \forall k_{3} \in K_{3} \label{equ.20_3}\\
& P_{t,k_{3}}+r_{t,k_{3}}^{cap,up} \le ER_{k_{3}}^{max};\hspace{3mm}\forall t\in T^{'},\, \forall k_{3} \in K_{3} \label{equ.20}\\
&  P_{t,k_{3}}-r_{t,k_{3}}^{cap,dn} \ge0;\hspace{3mm}\forall t\in T^{'},\, \forall k_{3}  \in K_{3}\label{equ.21}\\
\begin{split}\label{equ.23}
& 0.9CL_{k_{3}}^{max}b_{k_{3}}  \le E_{k_{3}}^{int}b_{k_{3}}+\sum_{t\in T^{'}}[ P_{t,k_{3}}+r_{t,k_{3}}^{cap,up} \mu_{t}^{up}\\
&\hspace{10mm}-r_{t,k_{3}}^{cap,dn} \mu_{t}^{dn}]  \gamma_{k_{3}}^{ch} \le CL_{k_{3}}^{max}b_{k_{3}} ;\hspace{1mm}\forall k_{3} \in K_{3}
\end{split}
\end{align}
where $T^{'}\subseteq T$ is the set of hours when EVs are available; $ER_{k_{3}}^{max}$ is the maximum charging rate; $ERR_{k_{3}}^{max}$ is the maximum permitted value for regulation capacity offers, $CL_{k_{3}}^{max}$ is the maximum charge level; $E_{k_{3}}^{int}$ is the initial charge level; $\gamma_{k_{3}}^{ch}$ is the charging efficiancy; $b_{k_{3}}$ is a binary variable which enables the DSO not to allocate the minimum power to EVCSs when their offering price is low. 

Equations (\ref{equ.20_1})-(\ref{equ.21}) limit EVCS's offers to the energy, regulation capacity-up and capacity-down markets. Equation (\ref{equ.23}) ensures that the charge level of EVs is full. 

\subsection{Constraints for Dispatchable DG Aggregators (DDGAGs)}
The operating constraints for DDDAGs are as follows:
\begin{align}
&P_{t,k_{4}}+r_{t,k_{4}}^{cap,up} \le P_{k_{4}}^{max};\hspace{3mm}\forall t\in T,\, \forall k_{4} \in K_{4} \label{equ.24}\\
&P_{t,k_{4}}-r_{t,k_{4}}^{cap,dn} \ge P_{k_{4}}^{min};\hspace{3mm}\forall t\in T,\, \forall k_{4} \in K_{4} \label{equ.25} \\
&0\le r_{t,k_{4}}^{cap,up} \le RU_{k_{4}} ;\hspace{3mm}\forall t\in T,\, \forall k_{4} \in K_{4}\label{equ.26}\\
&0\le r_{t,k_{4}}^{cap,dn} \le RD_{k_{4}} ;\hspace{3mm}\forall t\in T,\, \forall k_{4} \in K_{4}\label{equ.27} 
\end{align} 
where $P_{k_{4}}^{max}$ and  $P_{k_{4}}^{min}$ are the maximum and minimum power generations; $RU_{k_{4}}$ and $RD_{k_{4}}$ are the maximum ramp-up and ramp-down rates.

Equations (\ref{equ.24}) and (\ref{equ.25}) limit DDDAG's offers to the energy, regulation capacity-up and capacity-down markets. Equations (\ref{equ.26}) and (\ref{equ.27}) ensure the regulation capacity-up/capacity-down offers are lower than maximum ramp-up/ramp-down rates. 
\subsection{Distribution Power Flow Equations}
The linearized power flow equations are adopted from \cite{Baran}:
\begin{align}
\begin{split}\label{equ.28}
&\sum_{k_{1}\in K_{1}}\sum_{a\in A}H_{n,k_{1}}P_{a,t,k_{1}}+\sum_{k_{3}\in K_{3}}H_{n,k_{3}} P_{t,k_{3}}+P_{t,n}^{D}\\
&-\sum_{k_{2}\in K_{2}}H_{n,k_{2}} P_{t,k_{2}}-\sum_{k_{4}\in K_{4}}H_{n,k_{4}} P_{t,k_{4}}\\
&+H_{n}^{sub}P_{t}^{sub}+\sum_{j\in J}Pl_{j,t} A_{j,n} =0;\hspace{3mm}\forall t\in T, \, \forall n \in N \\
\end{split}\\
\begin{split}\label{equ.29}
&\sum_{k_{1}\in K_{1}}\sum_{a \in A}H_{n,k_{1}} P_{a,t,k_{1}} tan\phi_{k_{1}}+Q_{t,n}^{D}\\
& -\sum_{k_{4}\in K_{4}}H_{n,k_{4}} P_{t,k_{4}} tan\phi_{k_{4}}\\
&+H_{n}^{sub}Q_{t}^{sub}+\sum_{j\in J}Ql_{j,t} A_{j,n} =0 ;\hspace{3mm}\forall t\in T, \, \forall n \in N 
\end{split}\\
\begin{split}\label{equ.30}
& V_{m,t}=V_{n,t}-(r_{j} Pl_{j,t}+x_{j} Ql_{j,t} );\hspace{3mm}\forall t\in T,\,\forall m\in N,\\
&\hspace{10mm}\forall n \in N,\, C(m,n)=1 ,\, A(j,n)=1 \\
\end{split}\\
& V^{min} \le V_{n,t} \le V^{max} ;\hspace{3mm}\forall t\in T,\,\forall n \in N \label{equ.31}\\
& -Pl^{max} \le Pl_{j,t} \le Pl^{max};\hspace{3mm}\forall t\in T,\, \forall j\in J  \label{equ.32}\\
& -Ql^{max} \le Ql_{j,t} \le Ql^{max};\hspace{3mm}\forall t\in T,\, \forall j\in J  \label{equ.33}\\
\begin{split}\label{equ.35}
& r_{t}^{sub,up}=\sum_{k_{2}\in K_{2}}r_{t,k_{2}}^{cap,up}+\sum_{k_{4}\in K_{4}}r_{t,k_{4}}^{cap,up}\\
&\hspace{10mm}+\sum_{k_{1}\in K_{1}}r_{t,k_{1}}^{cap,dn}+\sum_{k_{3}\in K_{3}}r_{t,k_{3}}^{cap,dn};\hspace{3mm}\forall t\in T \\
\end{split}\\
\begin{split}\label{equ.36}
& r_{t}^{sub,dn}=\sum_{k_{2}\in K_{2}}r_{t,k_{2}}^{cap,dn}+\sum_{k_{4}\in K_{4}}r_{t,k_{4}}^{cap,dn}\\
&\hspace{10mm}+\sum_{k_{1}\in K_{1}}r_{t,k_{1}}^{cap,up}+\sum_{k_{3}\in K_{3}}r_{t,k_{3}}^{cap,up};\hspace{3mm}\forall t\in T  
\end{split}
\end{align}
where $H_{n,k}$ is the mapping matrix of DER aggregator $k$ to bus $n$; $P_{t,n}^{D}$ and $Q_{t,n}^{D}$ are the inelastic active and reactive power loads at each node; $Pl_{j,t}$ and $Ql_{j,t}$ are the active and reactive power flow at branch $j$; $A_{j,n}$ is the incidence matrix of branches and buses; $\phi$ is the phase angle; $C_{m,n}$ is the connecting nodes matrix.

Equations (\ref{equ.28})  and (\ref{equ.29}) represent the active and reactive power flow. Voltage drop at each line is represented by equation (\ref{equ.30}) and is limited by equation (\ref{equ.31}). Active and reactive power limits at each line are represented by (\ref{equ.32}) and (\ref{equ.33}).  Equations (\ref{equ.35}) and (\ref{equ.36}) represent DSO's aggregated offers for participating in the wholesale energy, regualtion capacity-up and capacity-down markets.
\section{Case Studies}
\begin{figure}
	\centering
	\includegraphics[width=0.95\columnwidth, height = 0.65in]{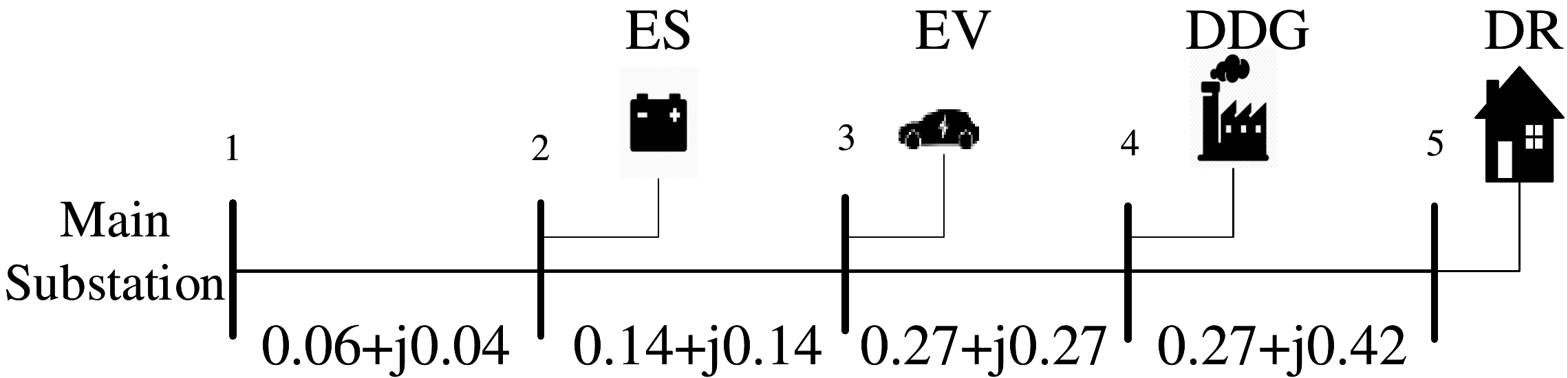}
	\caption{The small distribution network for case studies.}\label{fig.2.distribution network}
\end{figure}
\begin{table}
	\centering
	\caption{Market participants prices ($\$/MW$) and regulation signals ($P.U.$).}\label{table.1 input data}
	\begin{tabular}{p{0.1 cm}| p{0.3 cm} p{0.3 cm}| p{0.1 cm} p{0.1 cm}| p{0.2 cm} p{0.2 cm}| p{0.3 cm} p{0.3 cm} |p{0.2 cm} p{0.2 cm}| p{0.3 cm} p{0.3 cm}}
		\hline
		\multirow{2}{0.1cm}{t}&\multicolumn{2}{c}{Wholesale}&\multicolumn{2}{c}{ESAG}&\multicolumn{2}{c}{DDGAG}&\multicolumn{2}{c}{EVCS}&\multicolumn{2}{c}{DRAG}&\multicolumn{2}{c}{Regulation}\\
		\cline{2-13}
		&$E$&$C$&$E$&$C$&$E$&$C$&$E$&$C$&$E$&$C$&$up$&$dn$\\
		\hline
		1&24.3&14.7&25&23&28&27&29&30.5&29&30&0.45&0.42\\
	2	&23.7&17.3&25&23&28&27&29&30.5&29&30&0.45&0.42\\
	3	&23&16.6&25&23&28&27&29&30.5&29&30&0.45&0.42\\
	4	&23&16.6&25&23&28&27&29&30.5&29&30&0.45&0.42\\
	5	&23.7&17.3&25&23&28&27&29&30.5&29&30&0.45&0.42\\
	6	&25.9&22.7&28&25&29&28&29.5&31&30&31&0.48&0.48\\
	7	&29.4&30.4&28&25&29&28&29.5&31&30&31&0.48&0.48\\
	8	&30.7&33.6&28&25&29&28&29.5&31&30&31&0.48&0.48\\
	9	&30.1&33.6&28&25&29&28&29.5&31&30&31&0.48&0.48\\
	10	&29.1&31.4&28&25&29&28&29.5&31&30&31&0.48&0.48\\
	11	&28.8&30.4&28&25&29&28&29.5&31&30&31&0.48&0.48\\
	12	&28.2&24.3&28&25&29&28&29.5&31&30&31&0.48&0.48\\
	13	&27.5&24.3&27&24&28.5&27.5&29&30.5&29&30&0.5&0.51\\
	14	&27.2&24.3&27&24&28.5&27.5&29&30.5&29&30&0.5&0.51\\
	15	&27.2&24.3&27&24&28.5&27.5&29&30.5&29&30&0.5&0.51\\
	16	&27.5&24.3&27&24&28.5&27.5&29&30.5&29&30&0.5&0.51\\
	17	&28.2&28.2&30&27&29&28&29.5&31&30&31&0.5&0.51\\
	18	&30.4&28.8&30&27&29&28&29.5&31&30&31&0.5&0.51\\
	19	&32&33.6&30&27&29&28&29.5&31&30&31&0.5&0.51\\
	20	&32&33.6&30&27&29&28&29.5&31&30&31&0.5&0.5\\
	21	&31&32&30&27&29&28&29.5&31&30&31&0.5&0.5\\
	22	&29.4&32&28&25&29&28&29.5&31&30&31&0.5&0.5\\
	23	&27.5&25.6&28&25&28&27&29&30.5&29&30&0.42&0.45\\
	24	&25.3&22.4&28&25&28&27&29&30.5&29&30&0.42&0.45\\
		\hline
	\end{tabular}
\end{table}
Case studies are performed on the small distribution network shown in Fig.\ref{fig.2.distribution network}. The system contains $5$ nodes, where $N=\{1,2,3,4,5\}$; $4$ lines, where $J=\{1,2,3,4\}$; a DRAG, where $k_{1}=\{1\}$; an ESAG, where $k_{2}=\{2\}$; an EVCS, where $k_{3}=\{3\}$; a DDGAG, where $k_{4}=\{4\}$. The studies are performed over $24$ hours, $T=\{1,2,...,24\}$. EVs are available between hour $16$ and hour $24$, $T^{'}=\{16,17,...,24\}$. Initial charge level of ESAG is $8MW$. The following parameters are assumed: $\eta_{k_{2}}^{ch}=\eta_{k_{2}}^{di}=1$, $E_{k_{2}}^{min}=2MW$, $E_{k_{2}}^{max}=10MW$, $DR_{k_{2}}^{max}=CR_{k_{2}}^{max}=5MW$, $E_{k_{3}}^{int}=2MW$, $ER_{k_{3}}^{max}=5MW$, $ERR_{k_{3}}^{max}=0.5MW$, $P_{k_{4}}^{min}=0$,  $P_{k_{4}}^{max}=5MW$, $RU_{k_{4}}=RD_{k_{4}}=1MW$, $P_{a,t,k_{1}}^{max}=10MW$, $r_{k_{1}}^{cap,up,max}=r_{k_{1}}^{cap,dn,max}=1MW$.

The energy and regulation capacity prices in \cite{Fooladivanda} are considered. The hourly factors in \cite{Mousavi} are used to generate hourly prices. The regulation capacity-up and capacity-down prices are assumed to be equal. Regulation mileage-up and mileage-down prices are assumed to be equal. Regulation mileage prices are assumed to be $1/20$ of corresponding regulation capacity prices. Hourly energy prices, capacity up/down prices, and hourly regulation signals are given in Table \ref{table.1 input data}, where $E$ denotes energy price, $C$ denotes regulation capacity price.   

\begin{figure}
	\centering
	\subfigure[]
	{
		\includegraphics[width=0.35\textwidth, height=0.95in]{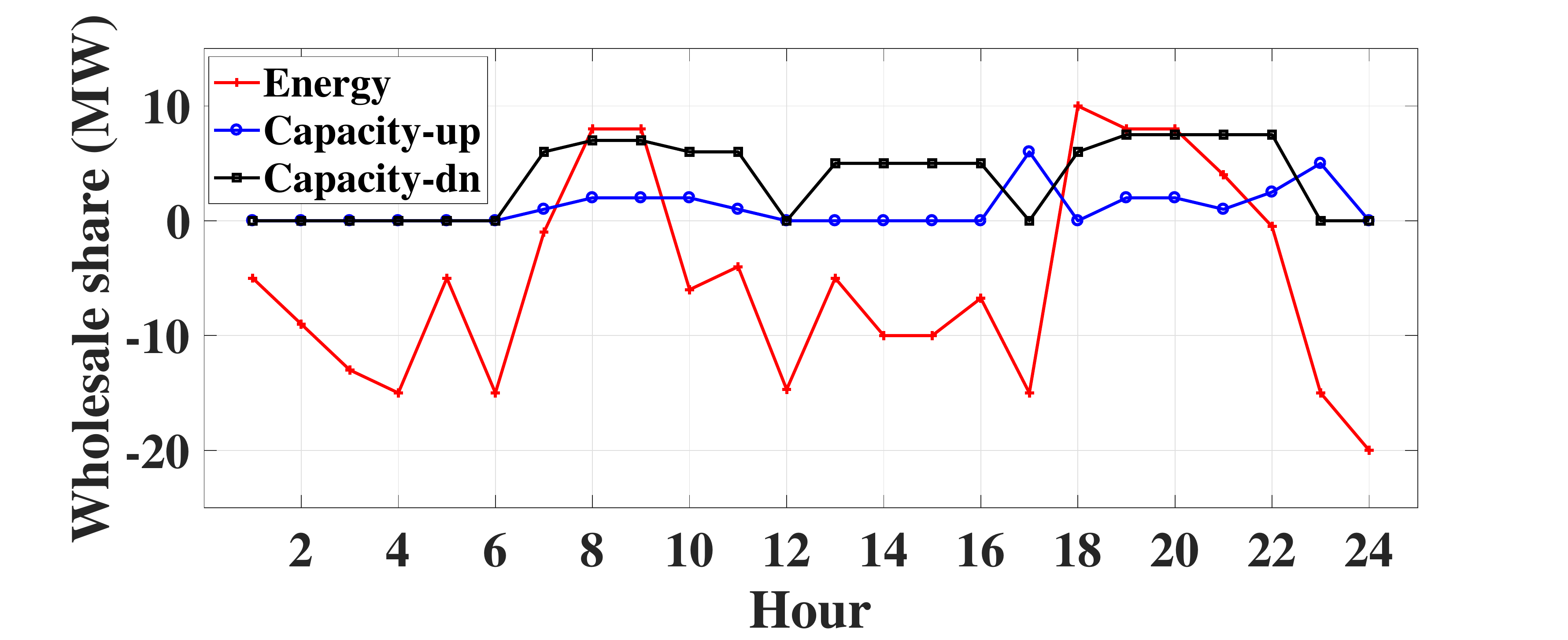}
		\label{fig.9.wholesalemarketshare}
	}
	\subfigure[]
	{
		\includegraphics[width=0.35\textwidth, height=0.95in]{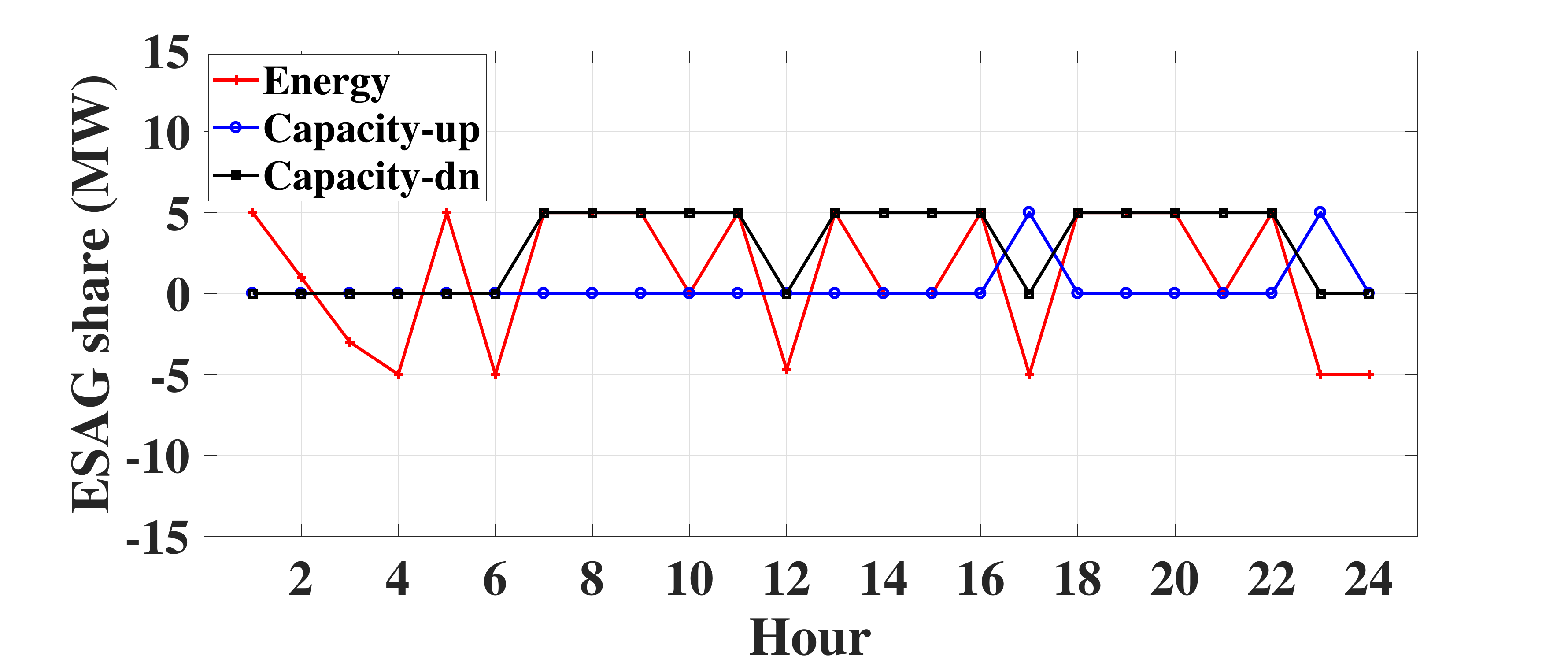}
		\label{fig.10.Esmarketshare}
	}
	\subfigure[]
	{
		\includegraphics[width=0.35\textwidth, height=0.95in]{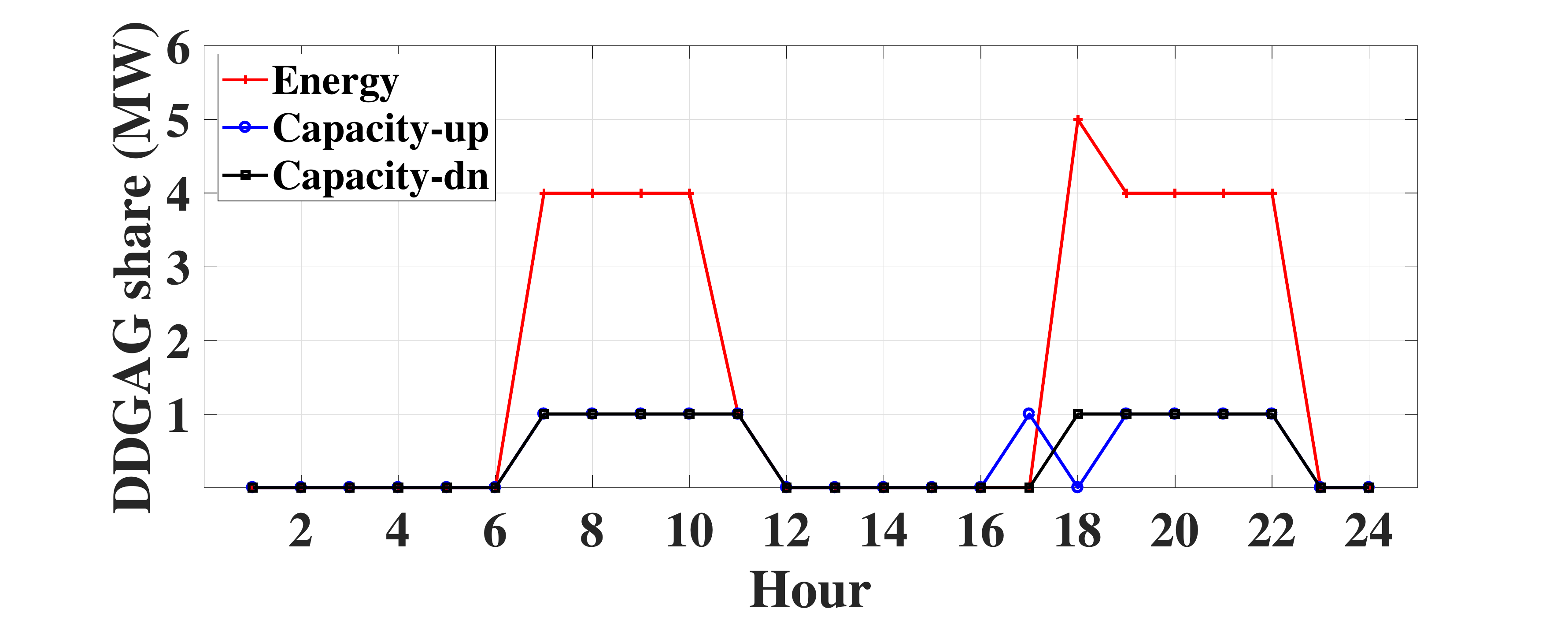}
		\label{fig.11.ddgmarketshare}
	}
	\subfigure[]
	{
		\includegraphics[width=0.35\textwidth, height=0.95in]{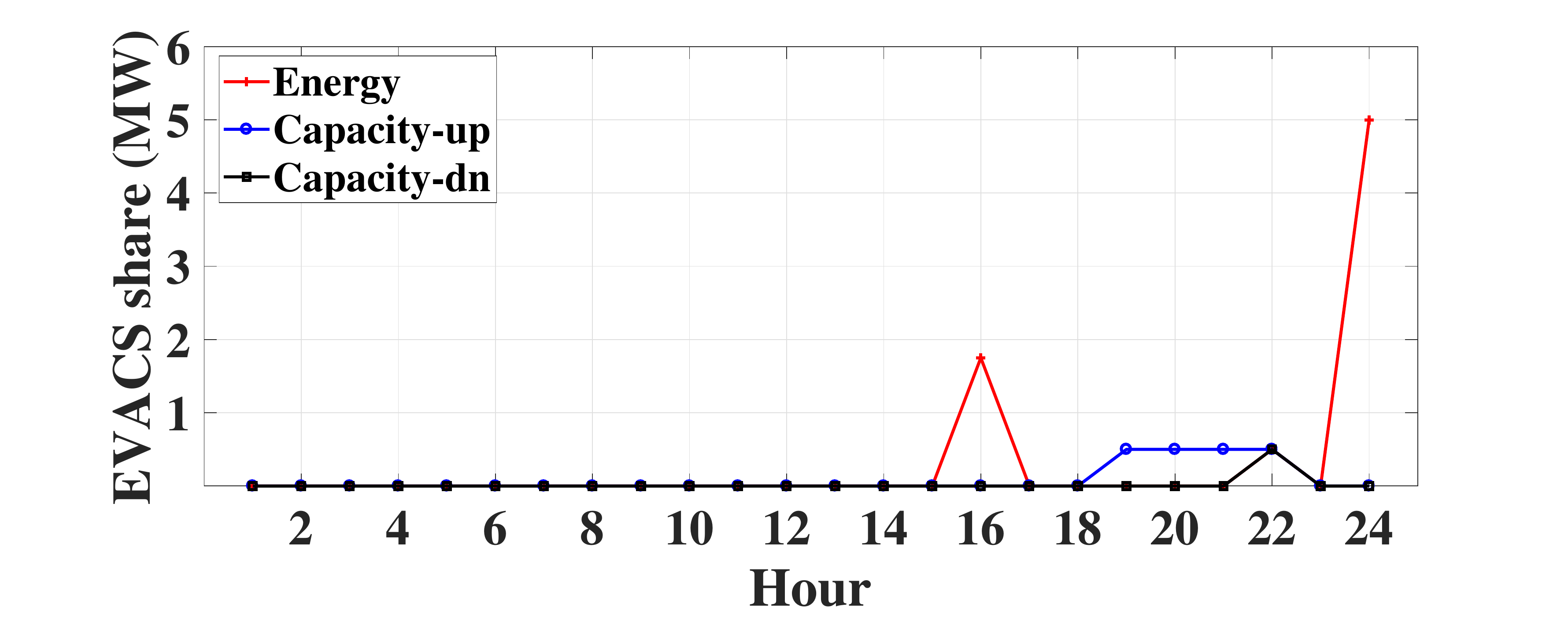}
		\label{fig.12.EVmarketshare}
	}
	\subfigure[]
	{
		\includegraphics[width=0.35\textwidth, height=0.95in]{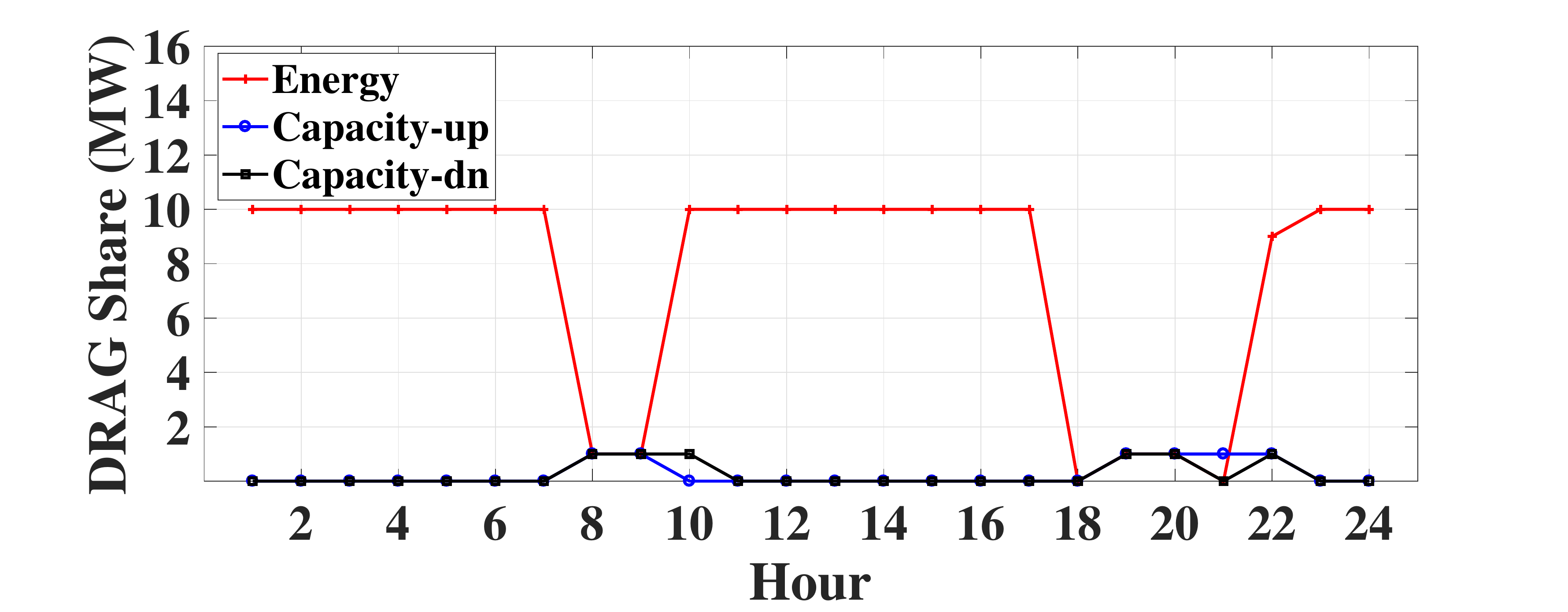}
		\label{fig.13.DRmarketshare}
	}
	
	\caption{Hourly awarded energy, regulation capacity-up/capacity-down services of (a) wholesale market (b) ESAG (c) DDGAG (d) EVCS (e) DRAG.}\label{Fig.3.marketoutcomes}
\end{figure}
\subsubsection{Market Outcomes}
The outcomes of DSO market coordination are presented in Fig. \ref{Fig.3.marketoutcomes}. The trades between the DSO and the wholesale market are shown in Fig. \ref{fig.9.wholesalemarketshare}. The awarded energy and regulation market shares of ESAG, DDGAG, EVCS, and DRAG are shown in Fig. \ref{fig.10.Esmarketshare}-Fig. \ref{fig.13.DRmarketshare}, respectively. At hours $ 8, 9,18, 19, 20, 21$, the DSO sells energy to the wholesale market since the prices of energy of the wholesale market at these hours are high. The DSO buys energy from the wholesale market at other hours.

The ESAG prefers offering regulation capacity-down service since this can increase its charging level. This causes the ESAG to offer regulation capacity-down service at hours $13, 14, 15, 16$, when the regulation capacity-dwon price is lower than the energy price in the wholesale market.

The DDGAG offers energy to the wholesale market at peak hours $7,8,9,10,11,18,19,20,21,22$. During these peak hours, the wholesale regulation capacity price is higher than the wholesale energy price. Hence, the DDGAG offers regulation capacity-up service at its maximum ramping rate ($1\, MW$). During peak hours, the DDGAG's remaining capacity ($4\, MW$) is offered to the wholesale energy market. However, at hour $18$, the DDGAG assigns all its capacity for energy provision, since the wholesale regulation capacity price is lower than the wholesale energy price at this moment.

The EVCS purchases energy at hours $16$ and $24$, when the wholesale energy price is the lowest among all the hours when EVs are available. The EVCS offers regulation capacity-up service at hours $19$-$22$, since 1) the wholesale regulation capacity-up price is high; and 2) the EVCS can increase EV charge levels by offering regulation capacity-up service.

The DRAG does not purchase energy from the wholesale market at peak hours. Also, it is not supplied by ESAG and DDGAG at peak hours, as they both sell energy to the wholesale market. However, the DRAG prefers offering regulation capacity to the wholesale market. Hence, it purchases energy that is enough for offering regulation capacity-down service.
\subsubsection{Sensitivity Analysis}
Sensitivity analysis is performed to study the impacts of ESAG's and DDAG's energy price offers on their revenue. For each study case $i \, (i=1,2,...,40)$, the market participants' energy price offers are modified from their base case values (in Table I) by a multiplier $i/10$.
\paragraph{ESAG Energy Price Offers}
\begin{figure}
	\centering
	\includegraphics[width=0.65\columnwidth, height=0.81in]{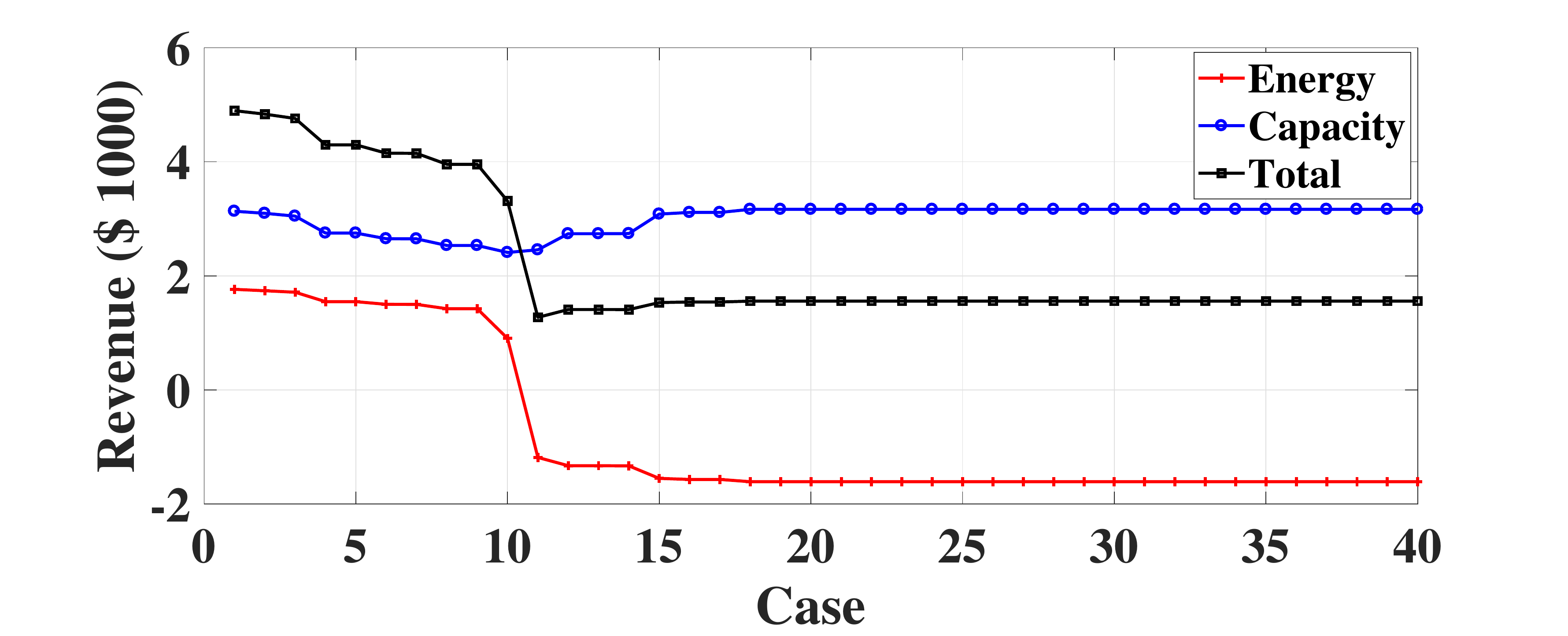}
	\caption{Variation of revenue of ESAG.}\label{fig.14.ESparameter}
\end{figure}
Fig. \ref{fig.14.ESparameter} shows the sensitivity of ESAG's total revenue with respect to its energy price offers. In Case $1$ with the lowest ESAG energy price offer, the ESAG offers regulation capacity-down service at all times even when its price offer for regulation capacity-down service is lower than the wholesale regulation capacity-down price. This is because ESAG can increase its charging level by providing regulation capacity-down service, and the energy gained during this charging period can be offered to the energy market. Hence, the ESAG gains the highest total revenue in this case. As the ESAG's energy price offer increases (from Case $2$ to Case $11$), its total revenue decreases. In Case $11$, the ESAG gains the lowest total revenue. This is beacuse in Case $11$, the ESAG's revenue from regulation capacity market is the lowest, as ESAG only offers regulation capacity service at peak hours when the wholesale regulation capacity price is high. After Case $11$, the ESAG's energy price offer is higher than the wholesale energy price. This causes the ESAG to act as demand and also offer regulation capacity-up service. By offering regulation capacity-up service, the ESAG decreases its charging level and increases its energy purchase from the energy market. Therefore, the ESAG's revenue from regulation capacity market increases after Case $11$ and becomes constant after Case $17$.
\paragraph{DDGAG Energy Price Offers}
\begin{figure}
	\centering
	\includegraphics[width=0.65\columnwidth, height=0.81in]{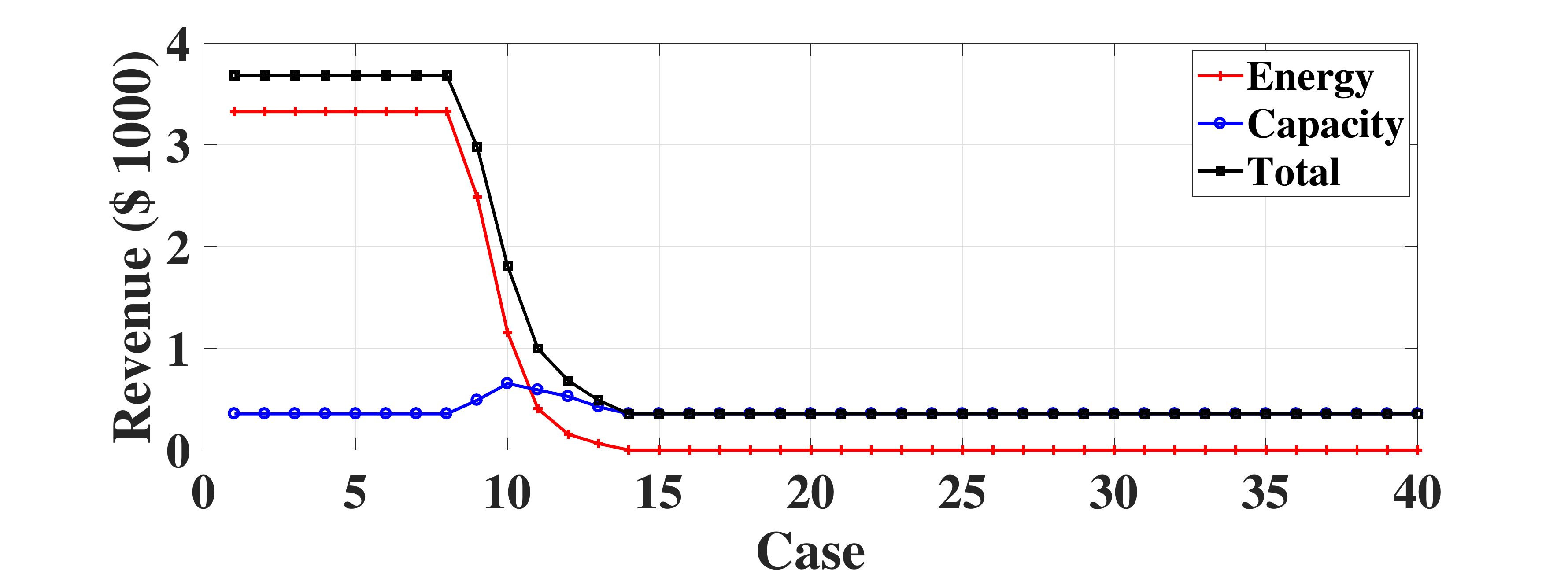}
	\caption{Variation of revenue of DDGAG.}\label{fig.15.DDGparameter}
\end{figure}  
Fig. \ref{fig.15.DDGparameter} shows the sensitivity of DDGAG's total revenue with respect to its energy price offers. Before Case $8$, the DDGAG's energy price offer is lower than the wholesale energy price at all the simulated hours. Hence, the DDGAG sells all the energy to the wholesale market while also providing regulation capacity-down service. In Cases $9$ and $10$, the DDGAG's energy price offer is lower than the wholesale energy price at some (not all the) simulated hours. Hence, it sells energy and provides capcaity-up service during these hours. This causes its energy revenue to decrease and regulation capacity revenue to increase. After Case $15$, the DDGAG's energy price offer is higher than the whole market price. This prevents the DDGAG from selling energy to the wholesale market, and also causes the DDGAG to provide regulation capacity-up service only. Therefore, the DDGAG's regulation capacity revenue becomes constant after Case $15$.
\section{Conclusion}
This paper proposes a DSO framework for coordinating DER aggregators to partcipate in the wholesale energy/regulation markets and retail energy market. Various types of aggregators are considered in the DSO operation. Case studies on a small distribution grid show the key interactions among wholesale energy/regulation markets, retail energy market operation, and DER aggregators' market participation. Sensitity analysis shows the DER aggregators' total revenue tends to decrease as they increase their energy price offers.


%
%

\bibliographystyle{IEEEtran}
\bibliography{References}

\end{document}